\documentclass[prd,preprint,superscriptaddress]{iopart}
\usepackage{graphicx}
\usepackage{iopams}
\newcommand{\be}{\begin{equation}}
\newcommand{\ee}{\end{equation}}
\newcommand{\ben}{\begin{eqnarray}}
\newcommand{\een}{\end{eqnarray}}
\begin{document}

\title{Coincidence problem within  dark energy as a coupled self-interacting Bose-Einstein gas}

\author{Jaime Besprosvany \footnote{
E-mail address: bespro@fisica.unam.mx}}
\address{ Instituto de F\'{\i}sica, Universidad Nacional Aut\'onoma de M\'exico, Apartado Postal 20-364, M\'exico 01000, D. F., M\'exico}

\author{Germ\'{a}n Izquierdo \footnote{ E-mail address:
gizquierdos@uaemex.mx}}\address{ Facultad de Ciencias, Universidad Aut\'onoma del Estado de M\'exico, Toluca 5000, Instituto literario 100, Edo. Mex.,M\'exico.}

\begin{abstract}

A  late accelerated expansion  of the  Universe is obtained from
non-relativistic particles with a short-range attractive
interaction,   and low enough temperature to produce a Bose-Einstein
condensate;  by considering coupled dark-energy particles,  energy is interchanged with dark matter, allowing it to describe recent acceleration by strengthening its effect. We show that for a sizable range of parameters,
dark energy and dark matter evolve with similar energy densities, solving
the coincidence problem, and in agreement with the luminosity distance vs redshift,  derived from supernova data.

\end{abstract}

\section{Introduction}

A  central tenet in modern cosmology is the existence of an element in its equations that leads to  universe acceleration.
Firmer evidence of the latter came with a more accurate measurement of the redshift vs distance, using distant   supernovae\cite{supernova}; additionally,    cosmic microwave background (CMB) anisotropy spectrum
\cite{wmap, planck}
and baryon acoustic oscillations (BAO)
\cite{baryonAcoustic}
   data  suggest  such a source becomes the more favored explanation. The implication is    the existence of a cosmological constant  or  dark-energy component within a general relativistic framework.

Indeed, the present model of cosmology has  two main
components: dark energy and dark matter, which account for most of the universe energy content. Although there is no laboratory evidence for their presence, indirect cosmological evidence has mounted. The model of universe most supported by observations is the $\Lambda$CDM model: a flat homogeneous metric with three main sources of energy at present: ordinary observable non-relativistic matter, cold dark matter (CDM) and dark energy (DE) in the form of a cosmological constant $\Lambda$ with constant energy density \cite{planck, reviews}.
CDM was necessary to explain motion within galaxies
\cite{galaxymotion}
and then galaxy formation. DE is also necessary in
structure formation,
\cite{structure}
  within   flat-space models, the latter required after evidence from the CMB, and  as implied by inflation.

A cosmological constant was first proposed in  early cosmology as a stabilizing element for the universe, while more conclusive evidence for its presence was obtained from supernova data\cite{supernova}. In the modern interpretation,  it is associated  to the energy-momentum part of the Einstein equations.
The $\Lambda$ term presents several problems, and
elucidating its  nature or that of dark energy and dark matter constitutes  an important objective.
A  relevant clue is derived from the work of Zeldovich\cite{Zel'dovich}, who associated the cosmological constant to a vacuum contribution, which produces such an acceleration. However, the associated scale of the vacuum, using either  Zeldovich's strong interaction scale\cite{Zel'dovich} or the fundamental Planck scale, constitute orders of magnitude beyond the present  scale of the universe energy density (and hence its components). This is known as the fine-tuning problem;  a natural way to solve it is to assume an evolving dark-energy component so that the connection between   the Planck scale and the value of the dark energy today can be explained. Bronstein\cite{Bronstein} first envisioned this kind of evolution.

The coincidence problem is closely related, and concerns   the unlikely similar  values of the dark-matter and dark-energy energy densities today; it may be resolved in models in which these components evolve similarly for long universe periods.

When the dynamical system presents an attractor, their energy-density ratio  may have reached its equilibrium value in the past, and we are no longer privileged observers as they will keep it  for a long time. Even if the attractor is in the far future, the coincidence problem can be alleviated as soon as the energy-density ratio evolves smoothly, as compared to the characteristic time span of the universe. A viable direction for the understanding of dark energy is to consider  models that reproduce this behavior, e. g., see Refs. in reviews \cite{CopelandSamiTsujikawa,Capozziello}. These models   address the coincidence problem successfully, but they do not have in general a description of the nature of the DE field; a degree of plausibility is gained with particle-physics models.

One alternative to the cosmological constant and the DE field, consist in the coupled dark energy models: the DE field interchanges energy with the CDM by means of a coupling term.
In this case, it is natural that both elements present a similar density at the present day, as the energies evolution of both are not independent. Several coupling terms can be found in the literature \cite{Yoo, PavonAtrioOlivares} some of them based in thermodynamics \cite{PereiraJesus, Besprosvany}, dimensional analysis \cite{Wang, Costa}, in quantum field theory \cite{Amendola, Zimdahl}, and particle physics \cite{Bento}.   We rely on the idea that these components are linked\cite{PadmanabhanChoudhury};   then, such a model is also consistent with an unification idea of CDM and DE. To obtain scaling we use a general  decay form present in many physical processes\cite{Barrow}.

In this work, we  rely  on the  DE model   presented in Ref.   \cite{interactBEG}, with a definite  microscopical nature, and show that it solves (or   alleviates) the coincidence problem.  The role of the DE field is played by a Bose-Einstein gas of non-relativistic particles that  self-interact attractively, which results in    a negative pressure. We also assume  gas particles are  created  at a defined rate, which enhances the acceleration  effect,  and results in a   coupling  between  the Bose gas and CDM. The dynamics of this model match well the observed data at  present, for some parameter  choices: the energy densities of both CDM and DE, the accelerated expansion, and the supernovae type Ia data. Additionally, different choices of the parameters lead to different luminosity distances at high redshift ($z>1.5$), giving us a tool to discard or validate the gas model in the near future. In  some coupled DE models\cite{PavonAtrioOlivares}, supernova acceleration is not  sufficiently constrained by the supernova data, but it is still necessary to check consistency. This we do in our paper.

This article's plan   is as follows. In section \ref{sec1}, we describe the Bose-Einstein gas with self-interaction in a cosmological frame considering the possibility that there is a coupling term between CDM and the gas energies. In section \ref{sec2}, we asume the rate of gas particle creation takes a feasible physical form, we deduce the coupling term from it and we solve the dynamics of the universe in terms of some free parameters; we also bind  the free parameter space by  imposing some observed results, and showing consistency  with  the coincidence-problem avoidance. In section \ref{sec3}, we illustrate the dynamics of the model for some choices of the parameters; we also address its future evolution . In section \ref{sec4}, we solve the evolution for a concrete set of parameters for which the coincidence problem is fully solved. In section \ref{sec5}, we summarize the findings of the previous sections.

From now on, we assume units for which $c=\hbar=k_B=1$. As usual, a zero subindex refers to the present value of the corresponding quantity; likewise, we normalize the metric scale factor by setting $a_{0} = 1$.

\section{Interacting Bose-Einstein gas particles and late acceleration}
\label{sec1}
\subsection{Energy density and pressure of an interacting Bose-Einstein gas}
We consider a self-interacting Bose-Einstein gas (IBEG) of non-relativistic particles that experiment a short-range two-particle attractive interaction ${\cal{V}}(x)$, thus, modifying free-particle behavior. In this section we summarize the description of the IBEG and we refer the reader to \cite{interactBEG} for details.

The average occupation number is
\ben
\label{occupation} \bar n_k=  \frac{1}{e^{(\epsilon_k-\mu)/T}-1 },
\een
where $\epsilon_k$ contains the kinetic energy of a particle
with momentum $\textbf{k}$ and mass $m$. The number of particles associated to a given energy state reads
\begin{eqnarray}\label{numberpar}
d N_k=  g \frac{V}{(2  \pi)^3} \bar n_k d^3 k,
\end{eqnarray}
where for spin-zero particles the degeneracy factor is $g=1$, and 
$V$ is the volume. The total number of particles is
\begin{eqnarray}\label{totalnumberpar} N=\int d N_k =N_c+N_{\epsilon} ,\end{eqnarray}
where $N_c$ and $N_{\epsilon}$ are the number of condensate and non-condensate particles, respectively.

At the critical temperature, $T_c= \frac{2 \pi}{\zeta(\frac{3}{2}  )^\frac{3}{2}}\frac{n^{2/3}}{g
^{2/3}m}\simeq  3.31 \frac{n^{2/3}}{g
^{2/3}m}$ (where the total particle number density is $n=N/V$, $m$ is the mass of the particle, and $\zeta$ is the zeta function), and also when $T< T_c$,  its single-particle energy given to first-order by \cite{Walecka}

\begin{eqnarray}\label {singleparticle}
\epsilon_{k}=m+\frac{k^2}{2 m}+ 2 v_0 ^\prime  n_{\epsilon}  ,
\end{eqnarray}
where the first term is the rest-mass energy, the second term is the kinetic energy, third term represents the potential energy of $N_{\epsilon}$ non-condensate particles interacting with the particle through the potential ${\cal{V}}(x)$, $n_{\epsilon}=N_{\epsilon}/V,$ and

\begin{eqnarray}\label {shortrange}
v_0^\prime=\int d^3x {\cal{V}}(x).
\end{eqnarray}
For  Eq. (\ref{singleparticle}), the corresponding chemical potential is

\begin{eqnarray}\label {chemicalpotenticalhighT}
\mu= 2 n_c v_0 ^\prime \label{mu}.\end{eqnarray} 
The constancy of $\mu$ remains  valid also in this case.

We can substitute the single-particle distribution  into the Bose distribution in Eq. (\ref{numberpar}) and integrate over the volume to find the energy,

\begin{eqnarray}\label
{energymodelsecond} E&=&mN+\frac{g V  m^{3/2}}{\sqrt{2} \pi^2} \int
d\epsilon
\epsilon^{3/2}\frac{1}{e^{(\epsilon_k-\mu)/T}-1} +\frac{1}{2V} \sum_{i\neq j}^{N} v_0  \\
&\simeq&mN+\frac{g V  m^{3/2}}{\sqrt{2} \pi^2} T^{5/2} \int_0^\infty
dz \frac{z^{3/2}}{e^{z}-1} + \frac{v_0}{2 V}N^2,
\label{en_gas}
\end{eqnarray}
where where $v_0 = v_0^
\prime[( N_c^ 2+2 (N-N_c)^2+2 N_c(N-N_c)]  / N^2$
is an interpolated potential term that takes into account the potential
particle exchange between the condensate and non-condensate
components, and the third term sums over pairs of interactions
from  Eq. (\ref{shortrange}) in the thermodynamic limit.

The total kinetic energy of the particles of the gas is related to its entropy, as the condensate particles do not contribute to the gas entropy. From the well-known equations for the Bose-Eintein gas \cite{Landau}
\[
E_c=\frac{3}{5}TS,
\]
\[
S=\frac{5}{3}(128g)^{2/5}m^{3/5}V^{2/5}E_c^{3/5},
\]
\[
E_c=\varepsilon T N_{\epsilon} \qquad ,
\]
where $\varepsilon=\frac{3 \zeta \left(\frac{5}{2}\right)}{2 \zeta \left(\frac{3}{2}\right)} \simeq 0.77$, it is possible to write the total kinetic energy as
\be E_c=A N_{\epsilon}^{5/3}V^{-2/3}, \ee
where
$A=\varepsilon^{5/3}(128g)^{-2/3}m^{-1}$. Thus, the kinetic energy density of the IBEG reads
\be \rho_c=A n_{\epsilon}^{5/3}. \label{kin_en_BEG} \ee

The total energy density of the IBEG reads from eqs. (\ref{en_gas}) and (\ref{kin_en_BEG})
\be \rho_g= m n+ A
n_{\epsilon}^{5/3}+\frac{1}{2}v_0 n^2, \label{endesBEG}\ee  and, from the definition $p=-\left(
\partial E /\partial V \right)_{N,S}$, the gas pressure takes the form

\be p=\frac{2}{3}\rho_c+\frac{1}{2}v_0n^2. \label{p_beg}\ee

In the next section, we consider the IBEG within a cosmological scenario.

\subsection{IBEG and late acceleration of the universe}

We assume a Lemaitre-Friedman-Robertson-Walker (LFRW) universe with line element

\[
ds^{2}=-dt^{2}+a(t)^{2}\left[ dr^{2}+r^{2}d\Omega_{\theta}
^{2}\right],
\]
where $a(t)$ is the scale factor and $t$, $r$ and $\Omega_{\theta}$ are the time, the radius and solid-angle comoving coordinates of the metric, respectively. As stated in Ref. \cite{interactBEG}, the above metric containing the IBEG with constant number of particles (with the horizon serving as boundary, making it stable despite the attractive interaction) cannot produce late acceleration. Indeed, the expansion of the universe becomes accelerated as soon as the interaction term dominates the expansion. However, as the reference volume evolves as $a^3$, this term scales as $a^{-6}$, while the kinetic and mass terms scale as $a^{-5}$ and $a^{-3}$, respectively. Consequently, a LFRW universe containing the IBEG expands with positive acceleration in its early stage but, eventually, it is dominated by the kinetic and mass terms, which leads to a decelerated expansion later.

We consider now a LFRW universe that contains baryonic matter,
cold dark matter (CDM) and the IBEG. The baryonic and CDM are both non-relativistic matter with equation of state $p=0$. Their energy densities ($\rho_b$ and $\rho_m$, respectively) evolve with the scale factor as $\rho_i=\rho_{i0} a^{-3}$ in the absence of coupling terms, where $i=b,m$, and $\rho_{i0}$ is the present-day density value.

If there is a particle creation process for IBEG, the particle number term is
no longer constant with the expansion. In fact, this creation mechanism produces an energy exchange between the CDM and the IBEG, described by a coupling term $Q$. The Einstein equations then read

\be H^2 = \left( \frac{\dot{a} }{a}\right)^2 =\frac{8 \pi G}{3}\rho_T, \label{eqH} \ee

\be \dot{\rho_b} + 3H\rho_b = 0, \label{evdensb}\ee
\be \dot{\rho_m} + 3H\rho_m = -Q, \label{evdensm}\ee
\be \dot{\rho_g} + 3H(\rho_g + p_g) = Q , \label{evdensg}\ee
where $\rho_T =\rho_b+\rho_m+\rho_g$ is the total energy density contained in the FLRW universe.

\section{Creation process of Interacting Bose-Einstein Gas particles }
\label{sec2}
A low temperature for the gas is chosen to enhance its dark-energy contribution. This temperature is maintained as the universe evolves \cite{interactBEG}. For simplicity, we assume that there is no condensation of IBEG particles. We also assume that the non-condensate IBEG particle number (with kinetic energy different from zero) evolves with time due to the coupling between CDM and the IBEG as a volume power law. Consequently,

\be N_{\epsilon}(t)=c V(t)^x,\label{N(V)}\ee
where $c>0$ and $1 \geq  x>0$ are the parameters that models this creation process from the CDM to the IBEG. This process is Markoffian, as it does not depend on previous states; typical dispersion processes and fluid interactions lead to evolution laws of this type in various physical setups \cite{Barrow}.

The IBEG non-condensate and condensate particle-number densities satisfy

\be \dot{n}_{\epsilon}+3Hn_{\epsilon}=3Hcxa^{3(x-1)},\ee
\be \dot{n}_c+3Hn_{c}=0, \ee
respectively, implying that they scale as
\be n_{\epsilon}=n_{\epsilon i}a^{-3}+ca^{3(x-1)},\label{ne}\ee
\be n_{c}=n_{ci}a^{-3},\ee
respectively.

In this work, we assume $x \leq 1$ in order to have a decreasing number density $n_{\epsilon}$ with the expansion of the universe. Allowing $x>1$, the model evolves to a scenario with both number particle and number density of IBEG increasing with time, making the interaction term eventually bigger than the kinetic and mass terms with a total energy of the IBEG particles negative.

We also assume that the IBEG creation is an ongoing process without initial conditions,
$n_{ci}=0$ and $n_{\epsilon i}=0$. Then, the total number of IBEG
particles at present is $n_0=n_{\epsilon}=c$.

With the above assumptions, the energy density of the IBEG $\rho_g$ reads from eq. (\ref{ne})

\be \rho_g = m c a^{3(x-1)}+ Ac^{5/3} a^{5(x-1)}+\frac{1}{2}v_0
c^2 a^{6(x-1)}. \label{endesBEG} \ee

By means of the energy conservation eq. (\ref{evdensg}), we can identify the coupling term $Q$ as

\be Q=3Hx \left(\rho_{G0}\,{a}^{3x-3}+\frac{5}{3} \rho_{c0}
a^{5x-5}+2\rho_{i0}a^{6x-6} \right), \label{int2}\ee where we have defined the parameters

\be \rho_{G0}=m c, \qquad \rho_{c0}=Ac^{5/3},\qquad
\rho_{i0}=(v_0 c^2)/2 . \label{parameters}\ee
We note that the energy density flow, $Q$ of above, can take negative values for certain choices of the parameters, for an early scale factor $a$. When $Q<0$, the energy flows from the IBEG to the CDM. Although IBEG particles are created, the energy can flow in the reverse direction as the IBEG  negative energy density interaction term $\rho_{i0}$ may predominate. As in eq. \ref{mu}, a connection between a coupling term $Q$ and the chemical potential has been pointed out before \cite{PereiraJesus, Besprosvany}. 

\subsection{Energy density evolution}

Once we know the form of $Q$, given by eq. (\ref{int2}), we can solve the set of eqs. (\ref{eqH}-\ref{evdensg}) analytically. The CDM energy density $\rho_m$ reads

\ben \rho_m =&\rho_{m0}a^{-3}-
\rho_{G0}a^{3x-3}\label{denm}\\ \nonumber
&-\rho_{c0}a^{5x-5}\frac{5x}{-2+5x}-\rho_{i0}a^{6x-6}\frac
{2x}{-1+2x},\een
the baryon energy density evolves as \be \rho_b=\rho_{b0}a^{-3},\label{densb} \ee
and, together with $\rho_g$ in eq. (\ref{endesBEG}), the total energy density is then
\ben
\rho_T
=&\rho_{b0}a^{-3}+\rho_{m0}a^{-3}\\ \nonumber
&+\rho_{c0}a^{5x-5}\left(1-\frac{5x}{-2+5x}\right)+
\rho_{i0}a^{6x-6}\left(1-\frac {2x}{-1+2x}\right).\label{densT}\een

The free parameters in our model are six: $\rho_{b0}$, the current baryonic energy density; $\rho_{G0}$, the mass contribution to the IBEG energy density; $\rho_{c0}$, the kinetic contribution to the  IBEG energy density; $\rho_{i0}$, the interaction term of the IBEG energy density (negative term due to its attractive nature); $\rho_{m0}$, the energy density of the CDM particles; and $x$, the exponent parameter that models the creation rate.

\subsection{Bounds to the free parameters of the model}
Our purpose is to find general bounds on the free parameters of the LFRW universe containing the IBEG, assuming that this model reproduces the observed energy densities of its components, and an accelerated expansion. We also have in view the coincidence problem as, for some choice of the parameters, it is solved or, at least, alleviated. The case with $x=1$ is treated separately in section \ref{sec4}.

\begin{enumerate}

\item{Present day energy densities}

We know from different observations of the CMB spectrum, BAO and supernovae, that the energy density of baryonic matter accounts for $4\%$ of  the total energy density of the Universe ($\rho_{b0}/\rho_{T0}=\Omega_{b0}=0.04$), the CDM energy density is $24\%$ ($\rho_{m0}/\rho_{T0}=\Omega_{m0}=0.24$), and the $72\%$ left
corresponds to the dark energy ($\rho_{g0}/\rho_{T0}=\Omega_{g0}=0.72$
) \cite{wmap}. At this point, we fix the energy density scale so that $\rho_{T0}=1$ (i.e. $3 H^2_0/(8 \pi G)=1$).

The role of the dark energy in our model is played by the IBEG. Evaluating eq. (\ref{endesBEG}) and eq. (\ref{denm}) at present (i.e., $a=1$), and imposing the observed relations between the energy densities, we obtain

\ben \rho_{m0}-
\rho_{G0}-\rho_{c0}\frac{5x}{-2+5x}-\rho_{i0}\frac
{2x}{-1+2x}=\Omega_{m0},& \\
\rho_{G0}+\rho_{c0}+\rho_{i0}=\Omega_{g0}.&  \een
As the baryonic matter in our model does not interact with any other sources of energy, $\rho_{b0}=0.04$. Using these relations, it is possible to express the free parameters $\rho_{c0}$ and $\rho_{i0}$ in terms of $\rho_{m0}$, $\rho_{G0}$ and $x$ as

\ben \rho_{c0}=&(-2\rho_{m0}-3\rho_{G0}+2\Omega_{m0})x^{-1} \\ \nonumber
& +9\rho_{m0}-9\Omega_{m0}-4\Omega_{g0}
+10\left( -\rho_{m0}+\Omega_{m0}+\Omega_{g0} \right) x,\\
\rho_{i0}=&2(\rho_{m0}-\rho_{G0}-\Omega_{m0})x^{-1} \\ \nonumber
& -9\rho_{m0}+4\rho_{G0}+9\Omega_{m0}+5\Omega_{g0}
+10\left( \rho_{m0}-\Omega_{m0}-\Omega_{g0} \right) x.\een

The definitions in eq. (\ref{parameters}), $\rho_{c0}>0$ and $\rho_{i0}<0$ limit the free parameters $\rho_{G0}$, $\rho_{m0}$ and $x$; Fig. \ref{fig1} shows the bounds on the $\rho_{m0}$-$\rho_{G0}$ space from this condition as well as for those obtained in (ii) and (iii).

\begin{figure}[tbp]
\includegraphics*[scale=0.35]{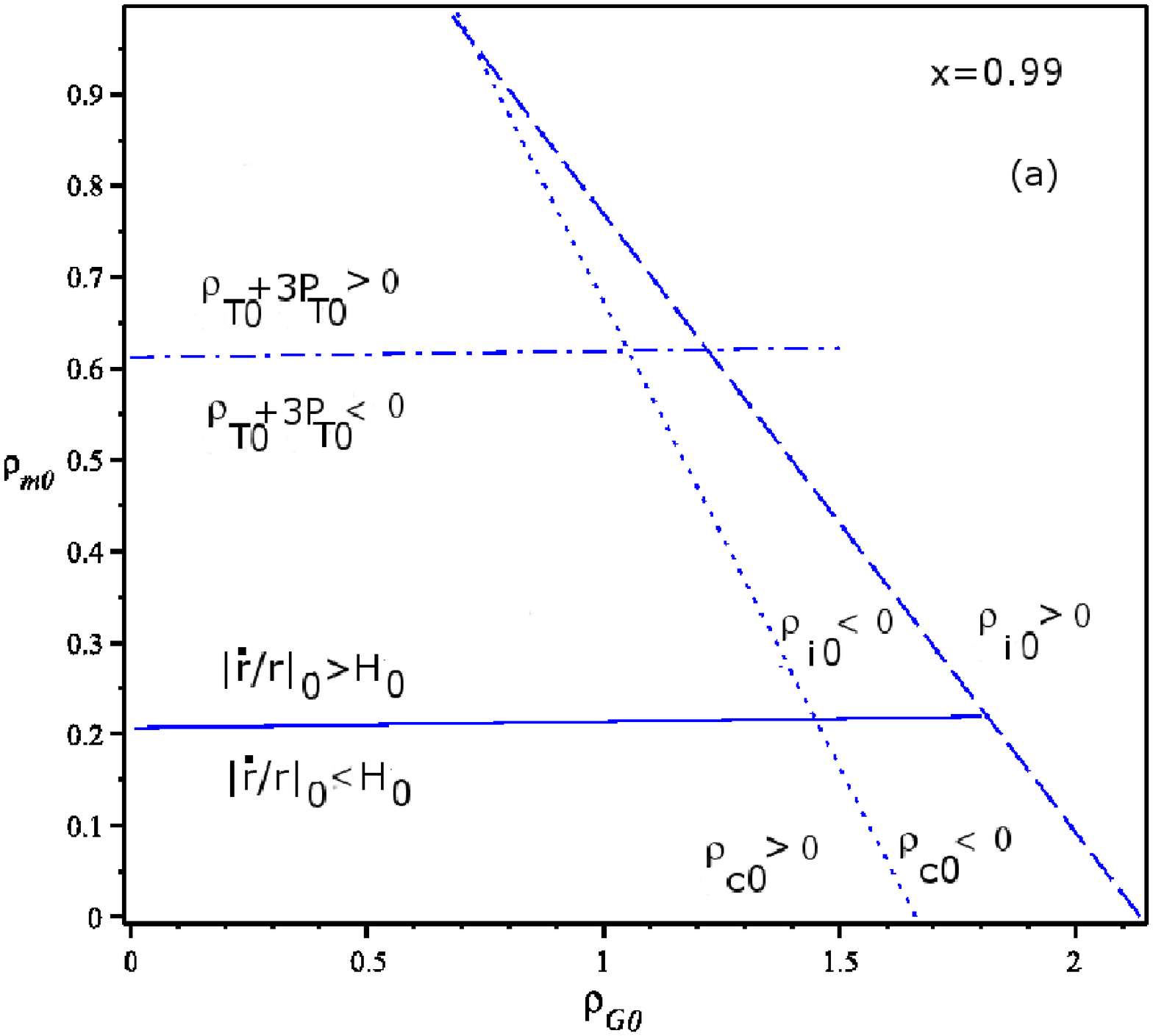}
\includegraphics*[scale=0.35]{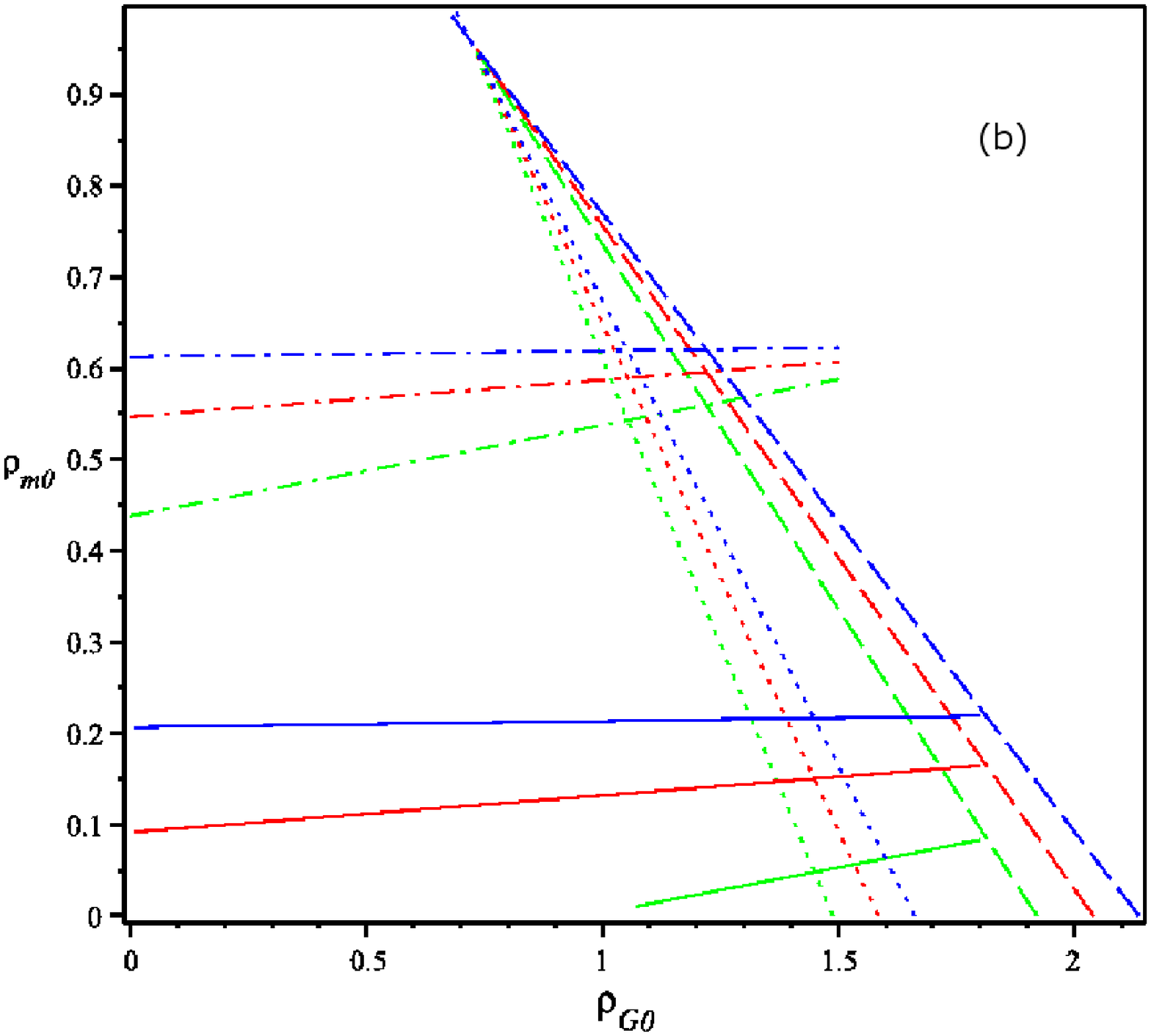}
\caption{(a): Bounds to the parameters for $x=0.99$. The dotted line represents condition $\rho_{c0}>0$, the dashed line represents condition $\rho_{i0}<0$, the dash-dotted line represents condition $\rho_{T0}+3p_{T0}<0$, and the solid line represents condition $|(\dot r /r)_0| \lesssim H_0$. (b): Bounds to the parameters for different $x$ values. The lighter line represents the $x=0.90$ case, the grey line represents the $x=0.95$ case, and the darker line represents the $x=0.99$ case.} \label{fig1}
\end{figure}

\item{Accelerated expansion of the universe}

As many observations suggest, the Universe is undergoing a
stage of accelerated expansion. In other words,
$\rho_{T0}+3p_{T0}<0$ (in order to obtain $\ddot{a}>0$ from the Einstein equations). From eqs. (\ref{p_beg}) and (\ref{densT}), this relation adds two new bounds to the parameters of our model: $x>1/2$, and
\ben
\rho_{m0} &<& \frac{0.04}{2-9x+10x^2}\left[250(\Omega_{m0}+\Omega_{g0})x^2\right. \\\nonumber
&&\left.-(250 \Omega_{m0}+200\Omega_{g0}+50\rho_{G0}+1)x+50\rho_{G0} + 50\Omega_{m0} \right]
;
\een

\item{Coincidence problem}

It is well known in the literature  that a dark-energy model solves the coincidence problem if the ratio $r=\rho_m/\rho_g$ tends to a constant \cite{Diegoetal}. In the IBEG model, $r$ does not evolve to a constant unless $x=1$ (see section \ref{sec4}). However, the coincidence problem is strongly alleviated when $|(\dot r /r)_0| \lesssim H_0$ \cite{Diegoetal}. If we impose this condition, we find an additional bound on the parameters.

From the definition of $r=\rho_m/\rho_g$, it follows that
\[
\frac{\dot{r}}{r}=\frac{\dot{\rho}_m}{\rho_m}-\frac{\dot{\rho}_g}{\rho_g}
\]
Derivating eqs. (\ref{endesBEG}), and (\ref{denm})  with respect to the time, we find the region of the $\rho_{m0}$-$\rho_{G0}$ space that fulfills the above condition. The region is represented in Fig. \ref{fig1} for fixed values of $x$.

For  $x\leq0.85$, our model of universe never fulfills the coincidence problem condition and conditions $\rho_{c0}>0$, $\rho_{i0}<0$ simultaneously. This is an additional bound on the parameter $x$ to the previous one ($x>1/2$ to obtain an accelerated expanding universe).
\end{enumerate}
The free-parameter space that fulfills all the above conditions for some  $x$ values is shown in Fig. \ref{fig2}. From the plot we can conclude that the closer the parameter $x$ is to $1$, the wider the $\rho_{m0}$-$\rho_{G0}$ region, i.e., the conditions imposed are less restrictive for values close to $x=1$.

\begin{figure}[tbp]
\includegraphics*[scale=0.35]{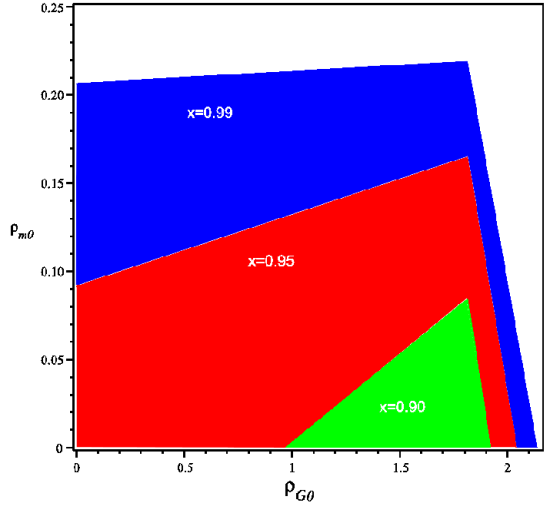}
\caption{Bounds on the free parameters.
 The region of the $\rho_{m0}$-$\rho_{G0}$ space that fulfills all conditions is colored in light grey for $x=0.90$, grey for $x=0.95$ and dark grey for $x=0.99$. Note that $x=0.99$ includes the $x=0.95$ region, and $x=0.95$ includes the $x=0.90$ region.} \label{fig2}
\end{figure}

In the next section, we study the evolution of  the different components.

\section{Numerical application and luminosity distance}
\label{sec3}
To study the evolution of the universe in the IBEG model, we look at their components evolution in eqs. (\ref{endesBEG}), (\ref{denm}) and (\ref{densb}). We plot the relative densities $\Omega_i=\rho_i/\rho_T$ as a function of the scale factor $a$ (with $i=g,m,b$). Next, we plot $log(r)$ vs. $a$ in order to check the coincidence problem in the model. Finally, we plot the effective magnitude $m_b$  vs. the redshift $z=1/a-1$ for each choice of the parameter together with the SNIa data from Ref. \cite{sncp}. The effective magnitude is defined as
\be m_B=5 log_{10}(d_L)+25,\ee
where $d_L$ is the luminosity distance. For the plot, we recover the units of the Hubble factor and take $H_0=75 km/s/Mpc$ \cite{wmap}.

\begin{figure}[tbp]
\includegraphics*[scale=0.25]{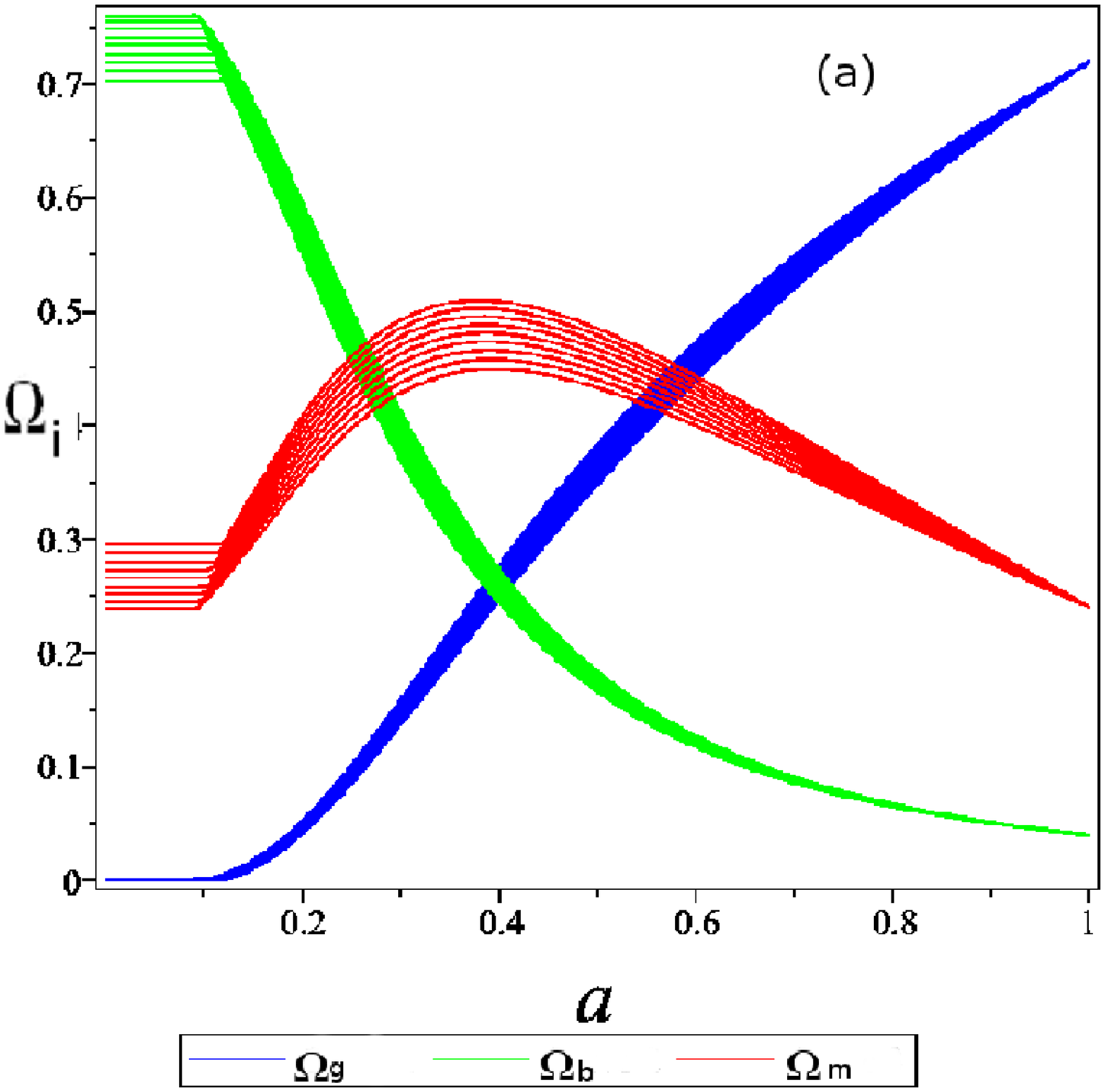}
\includegraphics*[scale=0.28]{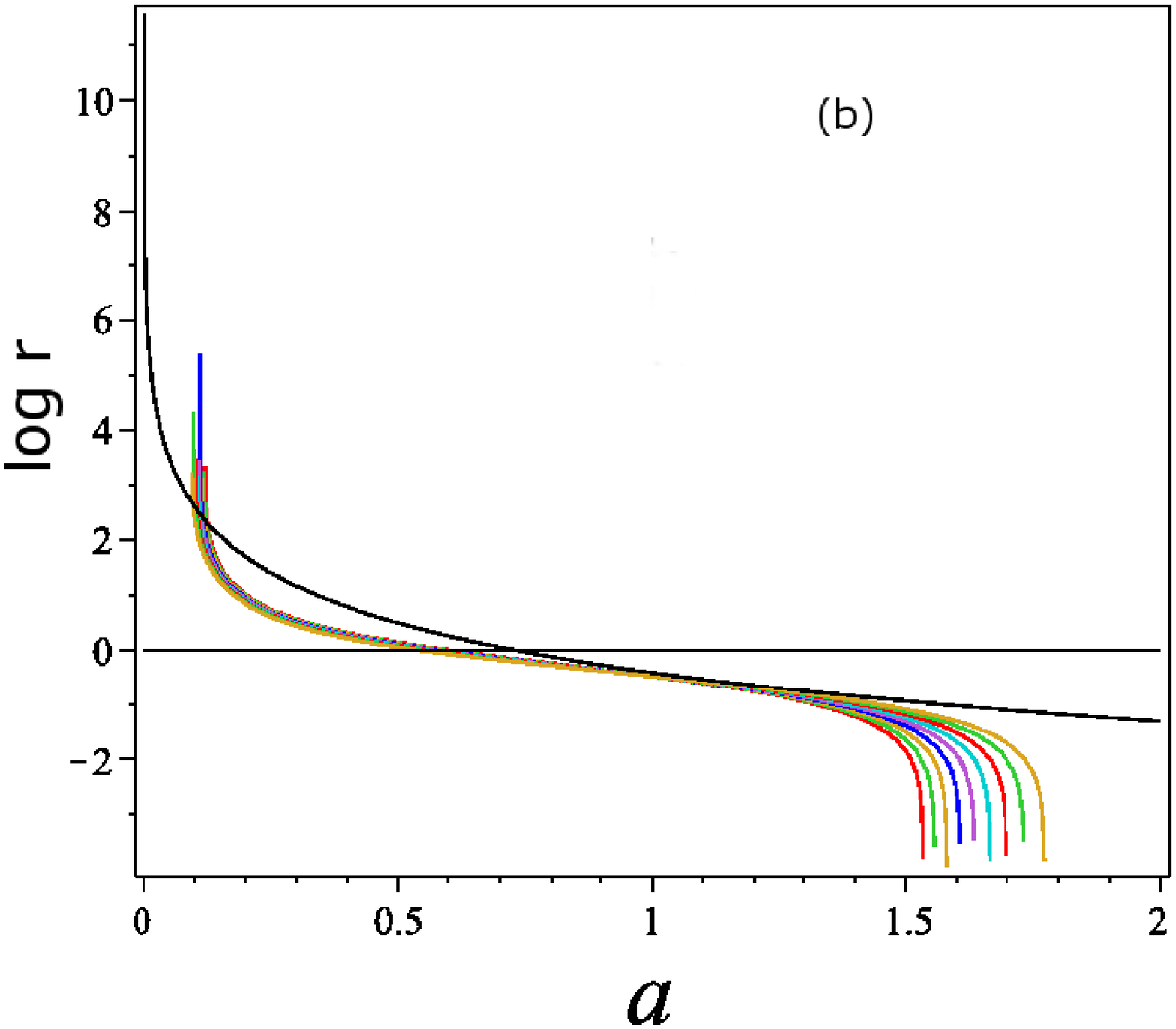}
\includegraphics*[scale=0.30]{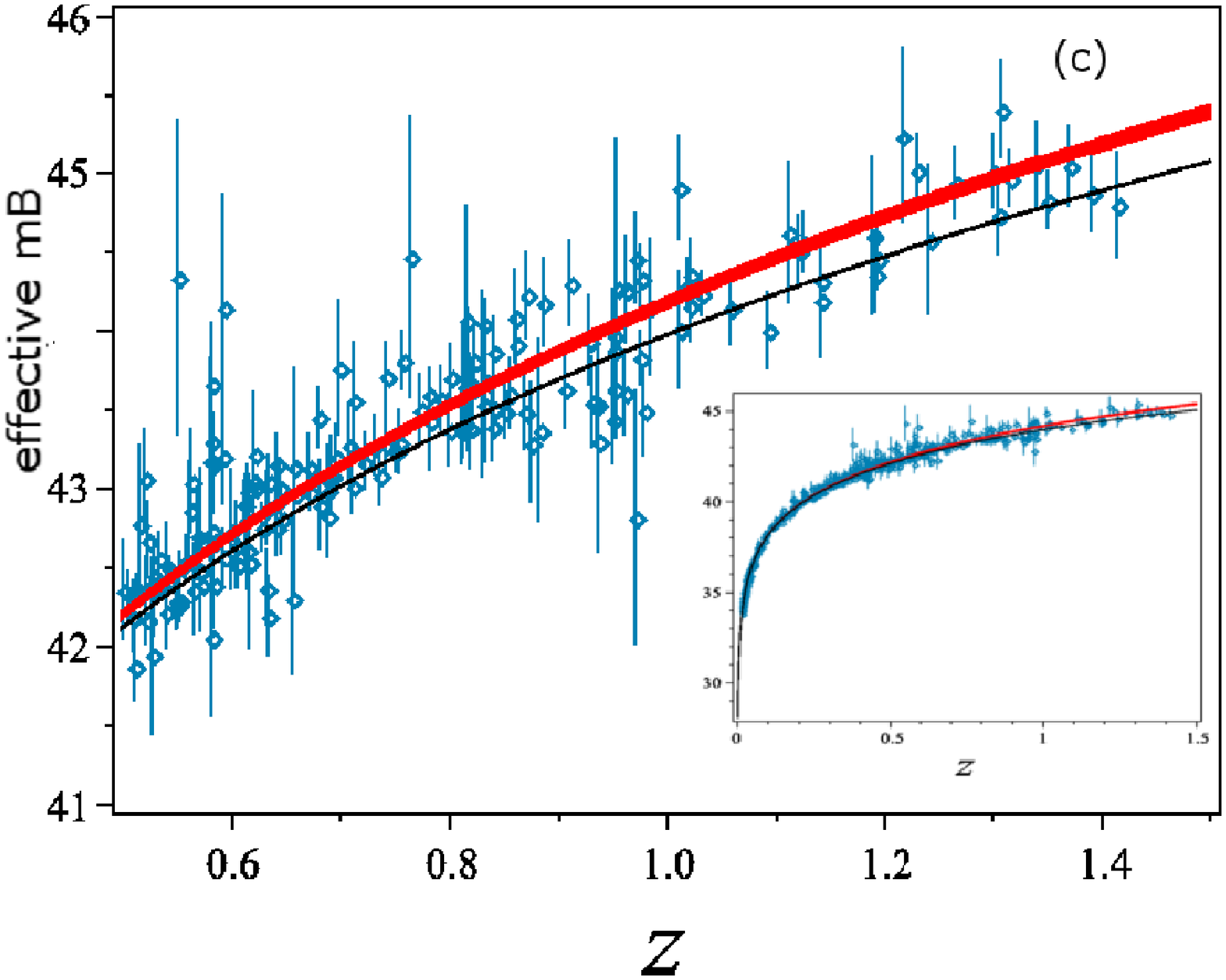}

\caption{IBEG model with $x=0.90$,
$\rho_{m0}=0.01$ and $\rho_{G0}\in[1.1,1.9]$ in steps of $0.1$. (a): Partial energy densities $\Omega_g$, $\Omega_b$ and $\Omega_m$ vs. scale factor $a$. (b): $log(r)$ vs. scale factor. (c): Luminosity distance vs. redshift ana observational data from supernovae type Ia \cite{sncp}. In the Panels b and c the solid black line represents the $\Lambda$CDM model. Please refer to the text for a detailed description of the plots.} \label{fig3}
\end{figure}

\begin{figure}[tbp]
\includegraphics*[scale=0.25]{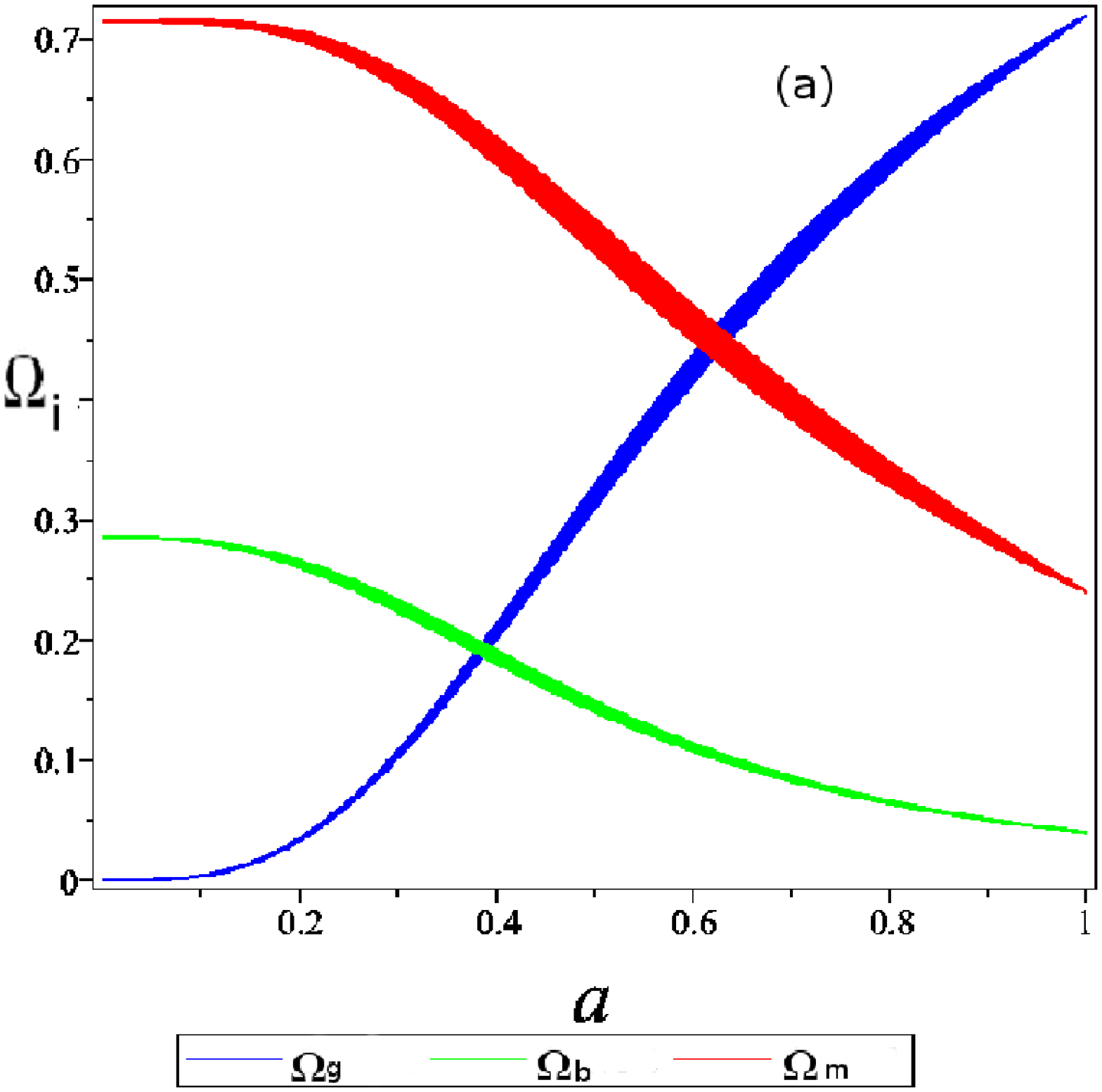}
\includegraphics*[scale=0.28]{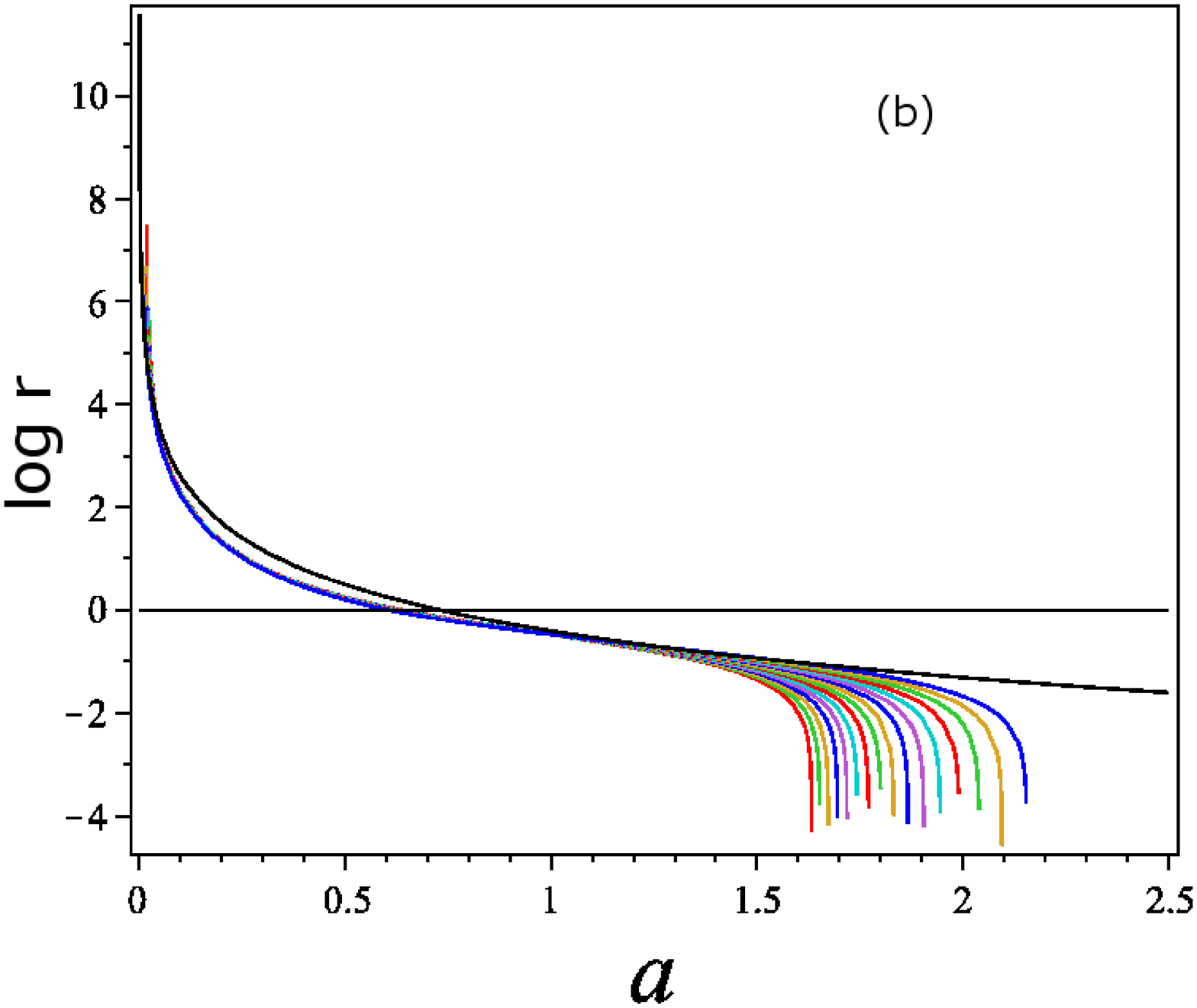}
\includegraphics*[scale=0.30]{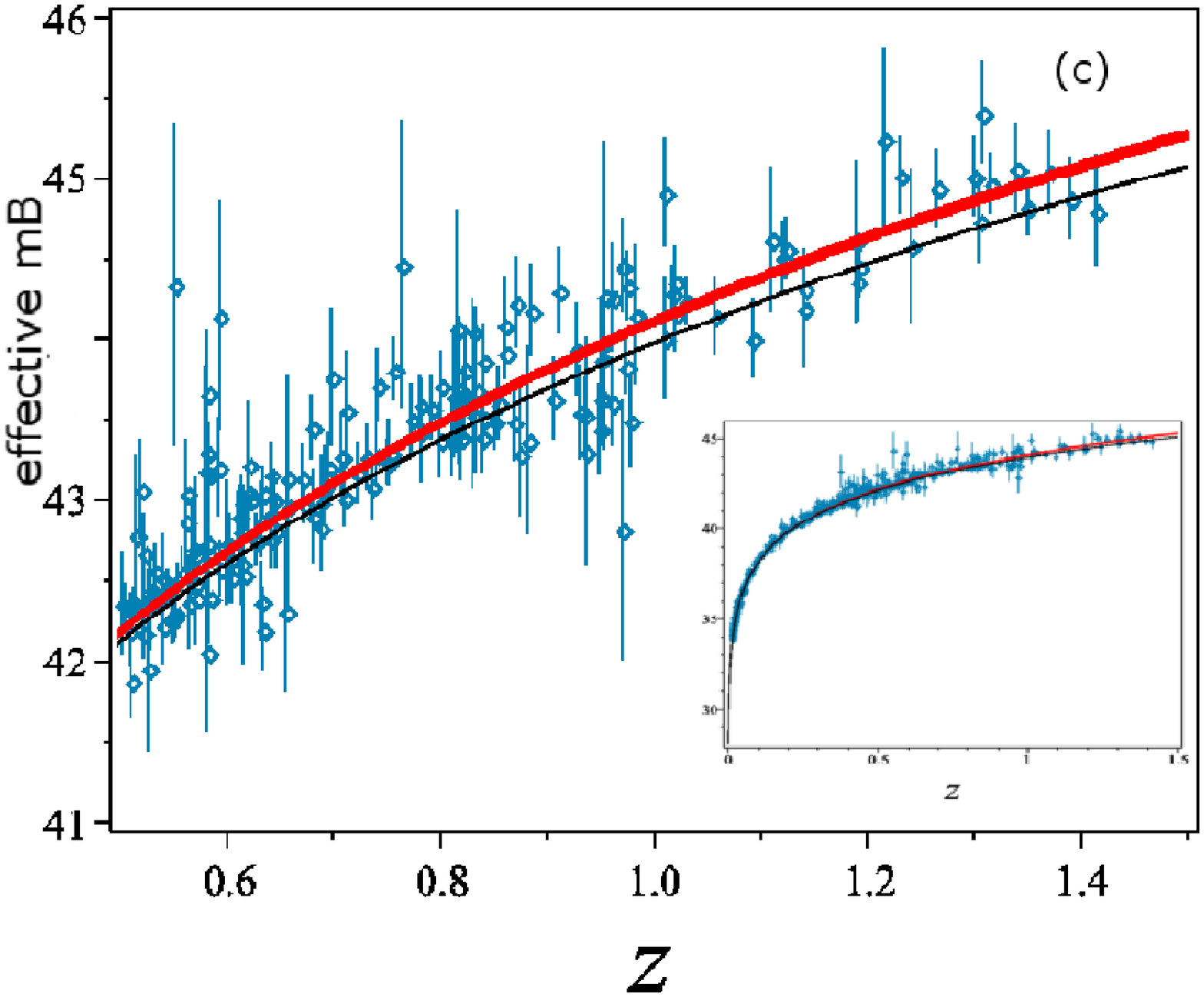}

\caption{IBEG model with $x=0.95$,
$\rho_{m0}=0.1$ and $\rho_{G0}\in[0.4,1.9]$ in steps of $0.1$. (a): Partial energy densities $\Omega_g$, $\Omega_b$ and $\Omega_m$ vs. scale factor $a$. (b): $log(r)$ vs. scale factor. (c): Luminosity distance vs. redshift and observational data from supernovae type Ia \cite{sncp}. In the Panels b and c the solid black line represents the $\Lambda$CDM model. Please refer to the text for a detailed description of the plots.} \label{fig4}
\end{figure}

\begin{figure}[tbp]
\includegraphics*[scale=0.25]{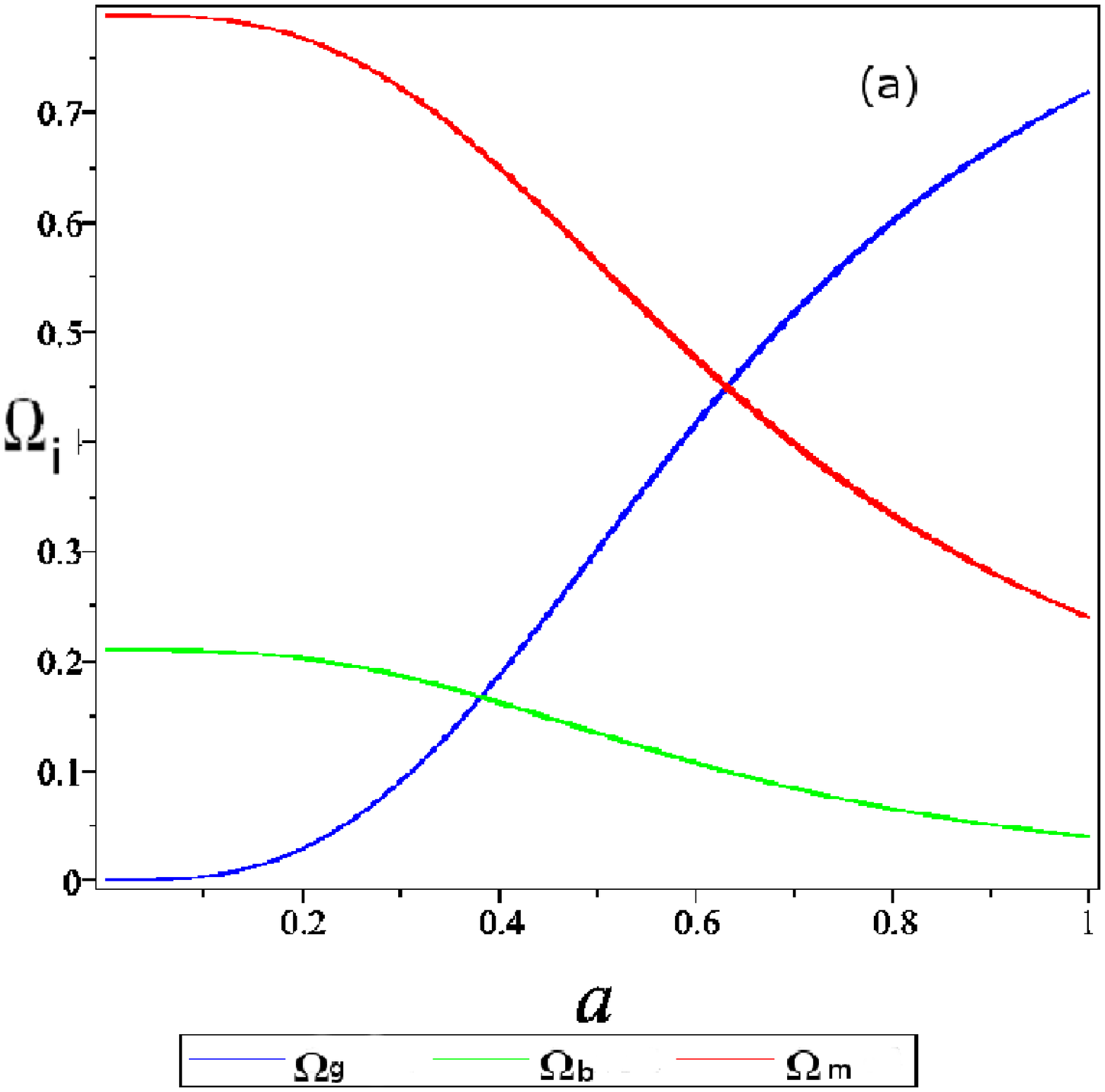}
\includegraphics*[scale=0.28]{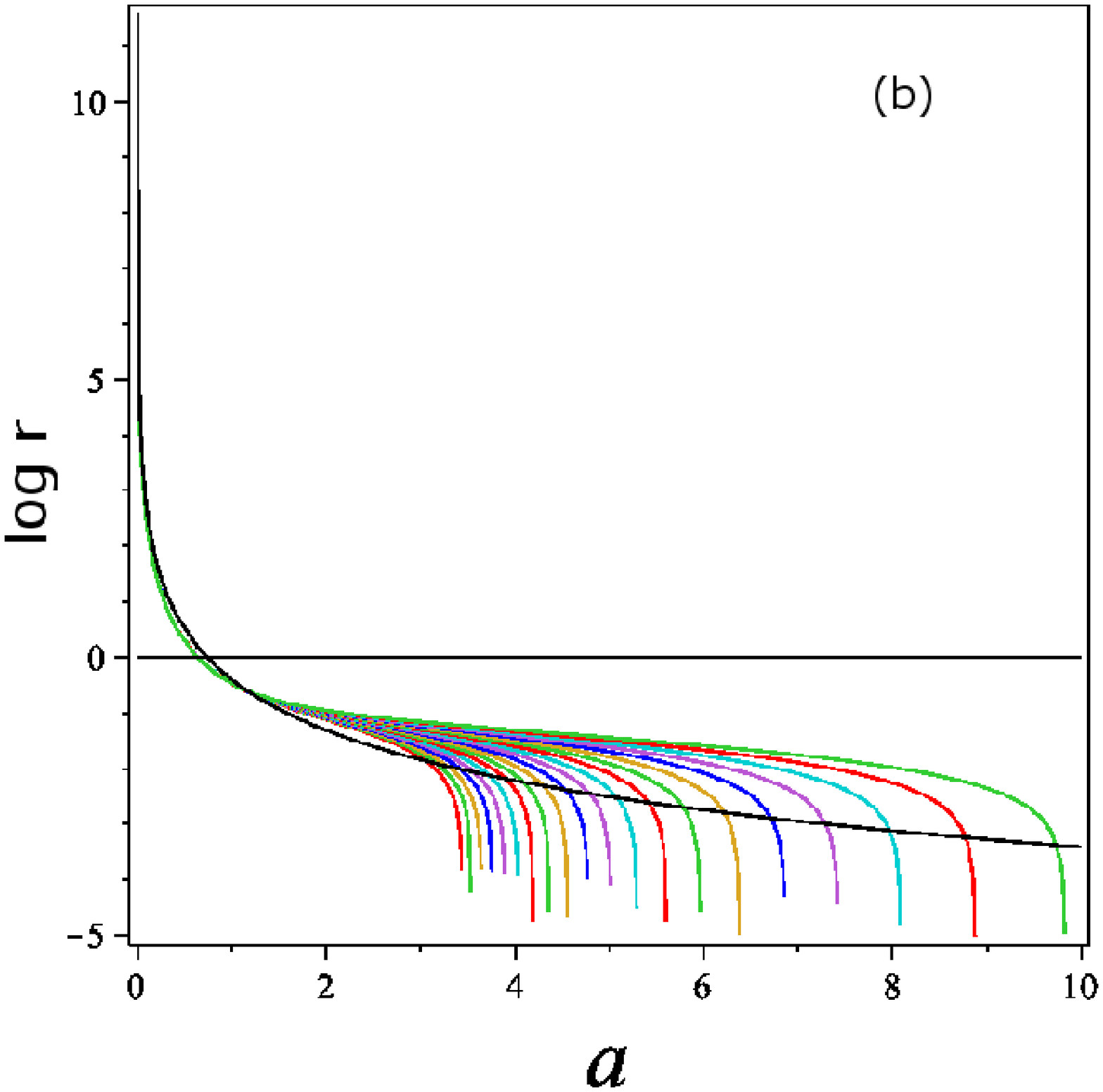}
\includegraphics*[scale=0.30]{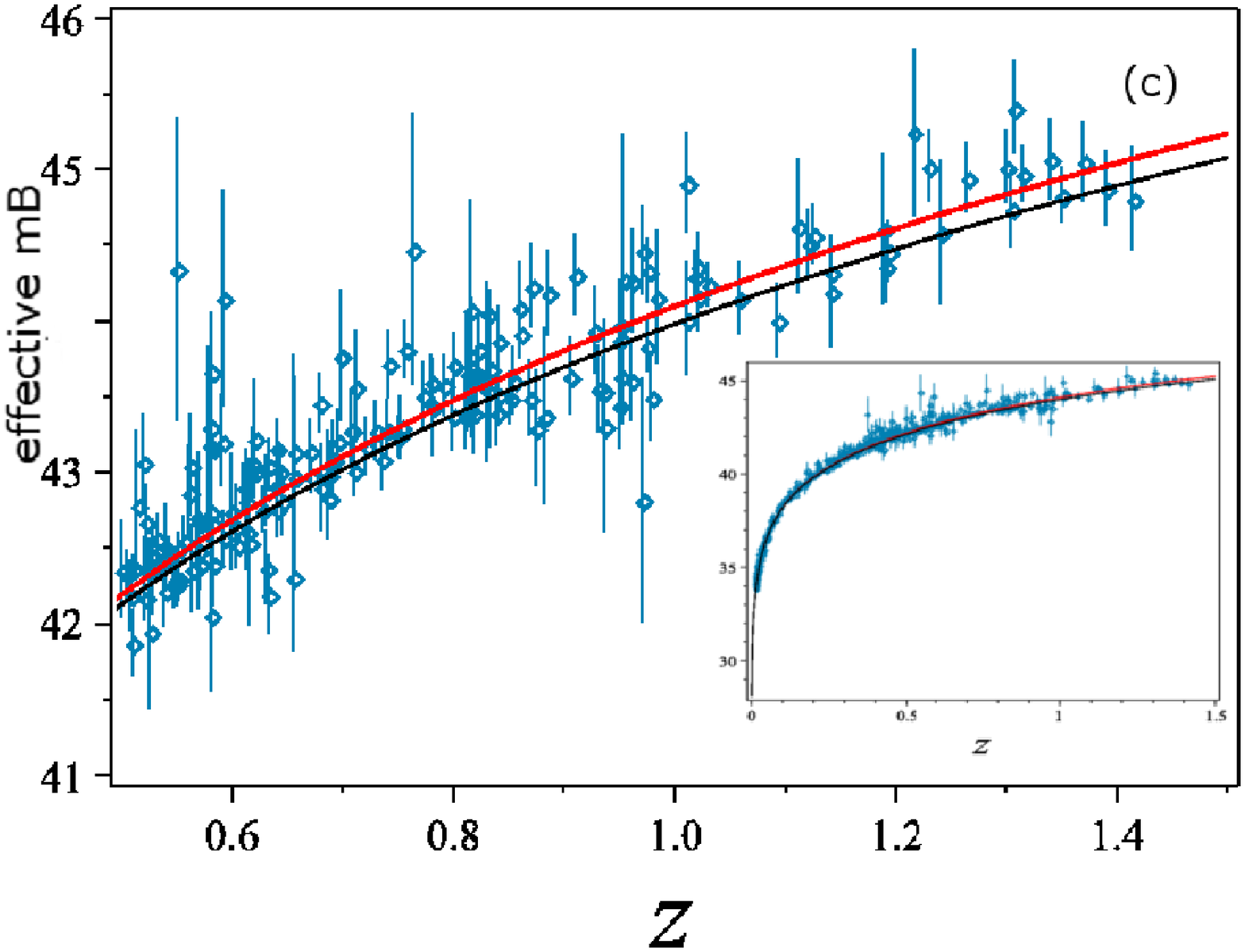}

\caption{IBEG model with  $x=0.99$,
$\rho_{m0}=0.15$ and $\rho_{G0}\in[0.0,1.9]$ in steps of $0.1$. (a): Partial energy densities $\Omega_g$, $\Omega_b$ and $\Omega_m$ vs. scale factor $a$. (b): $log(r)$ vs. scale factor. (c): Luminosity distance vs. redshift and observational data from supernovae type Ia \cite{sncp}. In the Panels b and c the solid black line represents the $\Lambda$CDM model. Please refer to the text for a detailed description of the plots.} \label{fig5}
\end{figure}

The plots for $x=0.90,0.95,0.99$ are given in Figs. \ref{fig3}-\ref{fig5}, respectively. The allowed region of parameters $\rho_{m0}$-$\rho_{G0}$ that fulfills the the previous section requirements is given in Fig. \ref{fig2}; we then fix the parameter $\rho_{m0}$, and we assign different values for the $\rho_{G0}$ parameter within the allowed region.

The plots for the different set of $\rho_{G0}$ and $\rho_{m0}$ show similar behavior. The $\Omega_i$ functions have a smooth dependence on $\rho_{G0}$ and $\rho_{m0}$.

Some common features are:
 \begin{itemize}
 \item The slope of the $r$ function is larger than that of the $\Lambda$CDM model until the creation process starts. At this point, it is smoothed out and the coincidence problem is alleviated with respect to the $\Lambda$CDM model in all the cases near $a=1$.

 \item The parameter $\rho_{G0}$ has a noticeable influence on the evolution of $\Omega_{g,m}$ and $r$ but not on the luminosity distance lines at the redshift range considered.

 \item The luminosity distance in the IBEG model is similar to that of the $\Lambda$CDM model in the redshift range $z < 0.5$. For the $z>0.5$ data, the IBEG model luminosity distance differs, while both models adjust the data well (both models have very similar chi-square statistics at $1\sigma$ confidence level). For larger values of $z$, the luminosity distance for the IBEG model depends on the choice of $x$ and is quite different from that of the $\Lambda$CDM model. Consequently, the IBEG model is distinguishable from the $\Lambda$CDM model and luminosity distance data from far away objects $z>1.5$ would discard/favor some values of $x$.
\end{itemize}
Some particular characteristics of each $x$ are:
\begin{itemize}
\item From Fig. \ref{fig2}, we see that the maximum allowed value for parameter $\rho_{m0}$ is $0.08$, $0.165$, and $0.219$ for $x=0.90,0.95,0.99$, respectively.

\item From Figs. \ref{fig3}a, and \ref{fig4}a, we conclude that the creation of IBEG process starts around $a=0.1$ for $x=0.90,0.95$ and all choices of the parameter $\rho_{G0}$, as $\Omega_g$ starts to grow from zero. From Fig. \ref{fig5}a, we obtain that the creation of IBEG process with $x=0.99$ starts very early (calculations show that as early as $a=10^{-8}$-$10^{-9}$) for all choices of the parameters.

\item A particular behavior of the $x=0.90$ case can be appreciated in Fig. \ref{fig3}a: $\Omega_m$ grows for a while after the IBEG creation process is triggered out. This fact is related to the interaction term proportional to $\rho_{i0}$ that appears in (\ref{denm}). This term is positive (as $\rho_{i0}<0$) and grows with the expansion. The evolution of the $\Omega_m$ is dominated by it for a while until the other decreasing terms start to dominate its evolution. This behavior of $\Omega_m$ is observed for all choices of the $\rho_{G0}$ and $\rho_{m0}$ parameters. In the $x=0.95, 0.99$ cases, $\Omega_m$ always decreases with $a$.

\item From Figs. \ref{fig3}b, \ref{fig4}b, and \ref{fig5}b,  we see that the creation process ends in the near future, when the CDM energy density and $r$ tend to $0$ drastically. This instant strongly depends on the choice of $\rho_{G0}$, varying from $a=1.5$ till $a=1.8$ in the $x=0.90$ case, up to $a=2.2$ in the $x=0.95$ case, and as far as $a=10$ in the $x=0.99$ case.
\end{itemize}

%
%\end{enumerate}

\subsection{Future evolution of the coupled IBEG model}

As stated in the numerical examples and Figs. \ref{fig3}-\ref{fig5}, the IBEG model predicts that the CDM energy density will
be consumed in the IBEG particle creation process at a future time. At this point, we assume the interaction $Q$ is set to $0$, to avoid a negative energy density of the CDM. The scale factor at this final moment depends on the parameters, in the previous section cases, it ranges from $1.5$ up to $10$, as can be appreciated in the evolution of $r$ in Fig. \ref{fig3}b-\ref{fig5}b.
At this point, the IBEG universe will be completely dominated by the IBEG energy density as the baryonic term will be much lower than it. The number density of IBEG particles will then evolve proportionally to $a^{-3}$, as no new particles will be created. The interaction term in the energy density $\rho_g$ is proportional
to $a^{-6}$  and will tend to 0 faster than the kinetic ($\approx
a^{-5}$) and the mass ($\approx a^{-3}$) terms. The accelerated
expansion stage will consequently end. Eventually, the kinetic term of the IBEG
term will also fade away and the mass term will remain. The
universe at this point will resume a second non-relativistic
matter-dominated era of expansion.

\section{Creation of Interacting Bose-Einstein Gas particles with $x=1$ }
\label{sec4}
The $x=1$ case must be resolved apart as it presents very different dynamics. When the number of IBEG particles is created at a rate such that $N_{\epsilon}(t)=c V(t)$, the number density remains constant with the evolution as

\be
n_{\epsilon}=c.\label{nex1} \ee
The energy density of the gas then reads

\be \rho_g = \rho_{G0}+ \rho_{c0}+\rho_{i0}. \label{endesBEGx1}
\ee
Thus, $\rho_g$ remains constant with the expansion of the universe. It plays a similar role to that of a cosmological constant in the $\Lambda$CDM model, except for the fact that there is a coupling term, and that the pressure is $p=(2/3)\rho_{c0}+\rho_{i0} \ne -\rho_g$.  Then,

\be Q=-3H \left(\rho_{G0}+\frac{5}{3} \rho_{c0} +2\rho_{i0}
\right). \label{int2x1}\ee

The CDM density $\rho_m$ from equation (\ref{evdensm}) is also different from that of the $\Lambda$CDM because of the coupling and it is

\be \rho_m =\rho_{m0}a^{-3}-
\rho_{G0}-\frac{5}{3}\rho_{c0}-2\rho_{i0}. \label{densmx1}\ee
The baryonic matter keeps its form as in the previous sections $\rho_b=\rho_{b0}a^{-3}$.

We still have the same free parameters as in the $x \neq 1$ case (except for $x$): $\rho_{m0}$, $
\rho_{G0}$, $\rho_{c0}$, $\rho_{i0}$ and $\rho_{b0}$. We can constrain the IBEG model's free parameters  with $x=1$ as in section \ref{sec2}, imposing the observed present energy densities, acceleration and coincidence-problem bounds.

\begin{figure}[tbp]
\includegraphics*[scale=0.35]{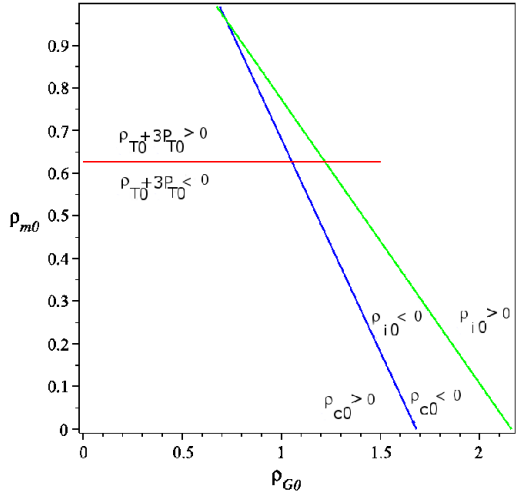}
\caption{Bounds on the free parameters for $x=1$. The region to the left of the darker line fulfills condition $\rho_{c0}>0$. The region to the left of the lighter line fulfills condition $\rho_{i0}<0$. The region under the grey line represents an accelerated expanding universe.} \label{fig6}
\end{figure}

\begin{enumerate}

\item Present-day energy densities

In this case, imposing $\Omega_{g0}=0.72$, $\Omega_{m0}=0.24$ and $\Omega_{b0}=0.04$ we
obtain the relations

\ben
\rho_{c0}&=&-3\rho_{G0}-3 \rho_{m0}+3\Omega_{m0}+6\Omega_{g0},\\
\rho_{i0}&=&2 \rho_{G0}+3\rho_{m0}-3\Omega_{m0}-5\Omega_{g0},\een
which allow us to reduce the free parameters space to a bi-dimensional $\rho_{m0}$-$\rho_{G0}$ space.
It is, again, possible to limit the two remaining free parameters by imposing
$\rho_{c0}>0$ and $\rho_{i0}<0$. The bounds to the $\rho_{m0}$-$\rho_{G0}$ region are represented in Fig. \ref{fig6}.

\item Present day acceleration

Another bound over the free parameters comes from requiring
that the expansion of the universe at present to be accelerated, i.e.
$\rho_{T0}-3p_{T0}<0$. This condition reads

\be \rho_{m0}<-0.01+0.67\Omega_{g0}+0.67\Omega_{m0}-3.33 \times 10^{-10}\rho_{G0}. \ee
In Fig. \ref{fig6}, the condition of above is represented as the region of $\rho_{m0}$-$\rho_{G0}$ under the grey line.

\item Coincidence problem

In this case, $r$ evolves to the constant
value

\be
r_{*}=\frac{\rho_{G0}-\frac{5}{3}\rho_{c0}-2\rho_{i0}}{\rho_{G0}+
\rho_{c0}+\rho_{i0}} .\ee Consequently, this model is
coincidence-problem free, and no extra bounds can be obtained from it.

\end{enumerate}

Fig. \ref{fig6} shows the region of the $\rho_{m0}$-$\rho_{G0}$ parameter space that fulfills the above conditions. The allowed region is less restrictive than that of the $x=0.90$,$ 0.95$, $0.99$ cases.

\begin{figure}[tbp]
\includegraphics*[scale=0.25]{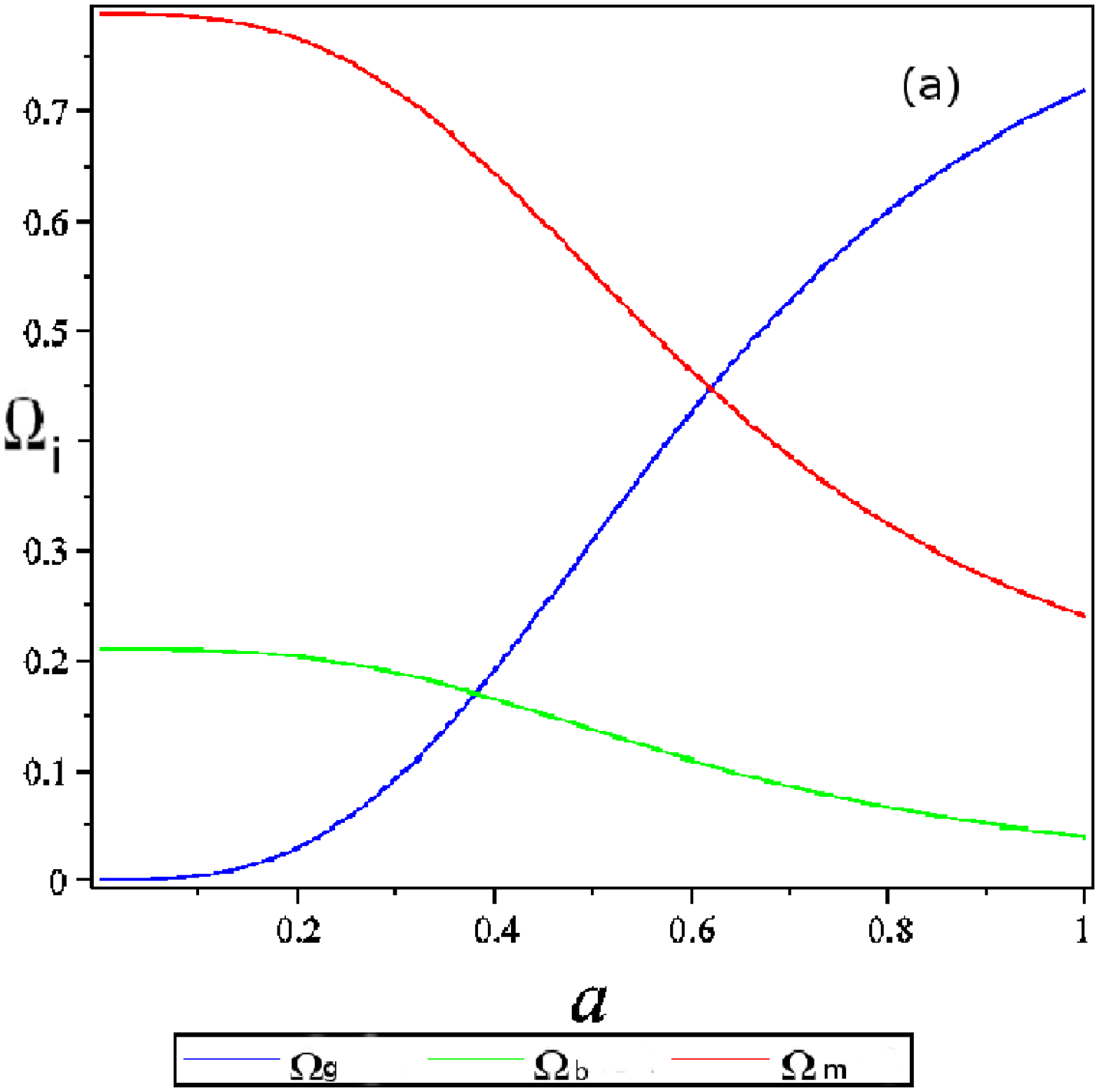}
\includegraphics*[scale=0.28]{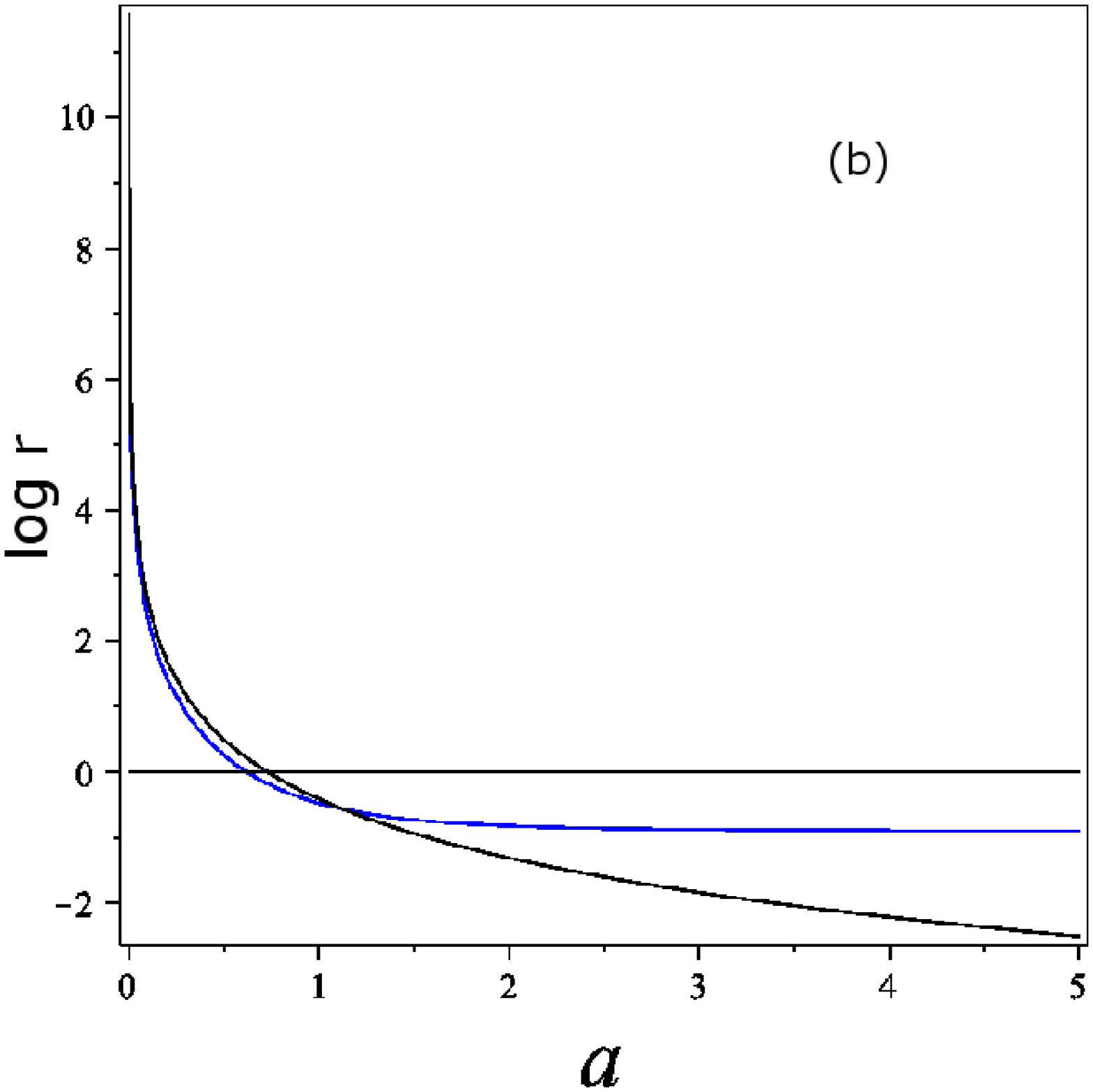}
\includegraphics*[scale=0.30]{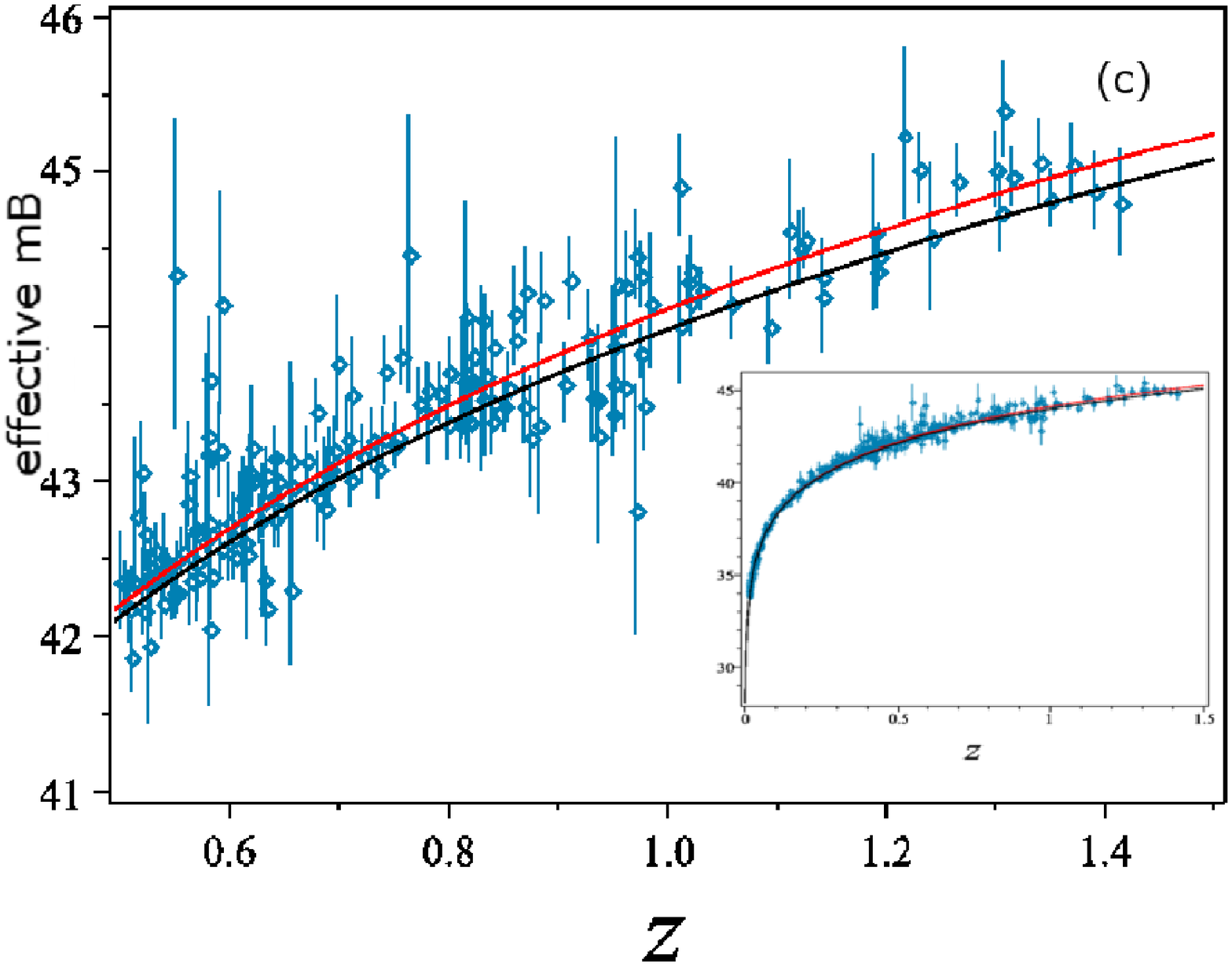}

\caption{IBEG model with  $x=1$,
$\rho_{m0}=0.15$ and $\rho_{G0}\in[0.0,1.2]$. (a): Partial energy densities $\Omega_g$, $\Omega_b$ and $\Omega_m$ vs. scale factor $a$. (b):  $log(r)$ vs. scale factor. (c): Luminosity distance vs. redshift and observational data from supernovae type Ia \cite{sncp}. In  (b) and (c)  the solid black line represents the $\Lambda$CDM model.} \label{fig7}
\end{figure}

In Fig. \ref{fig7}, we show the evolution of $\Omega_{b,g,m}$, $log(r)$ and the luminosity distance for the IBEG model with $x=1$. In this case, the maximum value of $\rho_{m0}$ allowed by the conditions is $0.626$. We show the findings of this case in figure \ref{fig7}. In Fig. \ref{fig7}a, we see that the energy density of CDM $\Omega_m$ is an always decreasing function of $a$. In this case, the creation process can start at any instant in the past, as the energy density of the IBEG remains constant. In Fig. \ref{fig7}b we observe that the slope of $r$ is smooth around $a=1$ and tends to a constant value for $a>1$. For this choice of parameters the creation process has no end. On Fig. \ref{fig7}c, we see again that this  choice of parameters is compatible with supernovae data. The IBEG model is closer to the the $\Lambda$CDM luminosity distance than the previous choices of $x$ for $z>1.5$, but it is still distinguishable.

The future evolution of the model in the $x=1$ case also differs
from the previous cases. If $\rho_{m0}<0.24$, the
creation process never consumes the CDM present on the universe and the accelerated expansion remains. In this case, the future evolution of the IBEG universe is dominated by a mixture of the IBEG (whose energy density acts as a cosmological constant, but with $p_g \ne -\rho_g $), and the CDM. From eq. \ref{densmx1}, it follows that, eventually, the CDM mass term proportional to $a^{-3}$ tends to zero and the remaining terms are also constants. The future evolution of the IBEG model with $x=1$ tends to a de Sitter universe.

On the other hand, if $\rho_{m0}>0.24$, the creation process will consume the CDM present in the future. The scale factor at this instant depends on the value of $\rho_{m0}$, but it is independent on the parameter $\rho_{G0}$. When the creation process ends, the IBEG universe will be dominated by the mass term $a^{-3}$ (as the interaction term will tend to zero faster), and the IBEG universe will evolve as a non relativistic matter-dominated universe.

\section{Conclusions}
\label{sec5}
In this work, we presented a physical model of DE: a self-interacting Bose-Einstein gas (IBEG), with an energy exchange with the cold dark matter (CDM) present in the universe that increases its particle number. A natural assumption is a connection between CDM and DE, and thus, it is plausible to consider such a coupling between the DE and the CDM. Indeed, our coupled DE model solves (or at least alleviates) the coincidence problem.

Assuming that our universe model contains baryonic matter, CDM and the IBEG, and that the creation mechanism only produces non-condensated IBEG particles at a general decay form $N=c V^x$ (a Markoff process), the dynamics of the universe follows and only six free parameters are left: $\rho_{b0}$, $\rho_{m0}$, $\rho_{G0}$ (the present day mass densities from baryonic, CDM and IBEG particles respectively), $\rho_{i0}$ (the energy density of the interaction term between IBEG particles), $\rho_{c0}$ and $x$ (both related with the creation mechanism). Additional bounds can be imposed over the free parameters. The first bound is obtained by imposing the present day energy densities values, $\Omega_{b0}$, $\Omega_{m0})$ and $\Omega_{g0}$ \cite{planck}, a second bound can be imposed by assuring the universe is undergoing an accelerated expansion stage at the present instant, i.e. $\rho_{T0}+3 p_{T0}<0$. Additionally, we can alleviate the coincidence problem in the IBEG model as soon as $|(\dot{r}/r)_0|<H_0$. These limits reduce the number of independent free parameters to 3: $\rho_{m0}$, $\rho_{G0}$ and $x>0.85$. If we fix additionally the value of the parameter $x$, the region of the $\rho_{G0}$-$\rho_{m0}$ space is bounded. Giving different values to the free parameters we observe that the creation process has a beginning at the instant $a_{in}$ determined by the choice of parameters. The IBEG model successfully fits the supernovae type Ia data with similar statistics than the $\Lambda$CDM model, and the parameter $\rho_{G0}$ has a low impact on the luminosity distance. The future evolution of the IBEG model presents a universe dominated by the IBEG particles acting as non-relativistic matter, once the CDM particles are exhausted and the self-interaction of IBEG particles has faded away.

If $x=1$, the dynamics of the IBEG model are quite different. First, the creation process has no starting point, the process can go in the past as far as the Big Bang. Secondly, the coincidence problem is solved, as the parameter $r$ tends to constant in the late expansion of the IBEG universe. In fact the IBEG presents a constant energy density with constant negative pressure (but $p \neq -\rho$ so in this sense we cannot affirm it acts as a cosmological constant). The luminosity distance in this case, adjusts the supernovae type Ia data with similar statistics to the $\Lambda$CDM. Also, when $\rho_{m0}<0.24$, the creation process never completely consumes the CMD present in the universe and the accelerated stage lasts forever similarly to a de Sitter universe.

In any case the IBEG model present a different luminosity distance at high redshifts ($z>0.5$) for every choice of parameters $x$ and $\rho_{m0}$, and also different than the $\Lambda$CDM one. If we obtain observational data from far away objects, we will be able to determinate for which set of parameters the IBEG model is more suitable in order to adjust them. Physical plausibility is gained by using microscopic models to predict cosmological variables  and, in turn, by using  cosmological data to constrain these models.

\end{document}